# High-resistance $YBa_2Cu_3O_{7-x}$ grain-boundary Josephson junctions fabricated by electromigration

M. Lyatti [1,2,*], U. Poppe [3], I. Gundareva [1,2] and R.E. Dunin-Borkowski [2]

[1] Kotelnikov IRE RAS, 125009 Moscow, Russia
[2] PGI-5, Forschungszentrum Jülich, 52425 Jülich, Germany
[3] CEOS GmbH, Englerstr. 28, 69126 Heidelberg, Germany

* E-mail: matvey_l@mail.ru

**Abstract.** [100]-tilt grain-boundary $YBa_2Cu_3O_{7-x}$ (YBCO) junctions are promising for investigation of macroscopic quantum phenomena in high-$T_c$ Josephson junctions. However, fabrication of the [100]-tilt grain-boundary YBCO junctions with a high resistance, which are required to study quantum effects, is difficult because of a high transparency of a tunnel barrier in this type of junctions. Here, we demonstrate a modification of grain-boundary barrier properties with a new approach to an oxygen electromigration in the YBCO grain-boundary junctions when the oxygen diffuses under an applied electric field from the grain-boundary to a $BaTbO_3$ layer deposited atop of an YBCO film. Using this approach, we changed the normal-state resistance of the junctions from tens to several hundred Ohms without a degradation of their characteristic voltage $I_cR_n$ and determined a barrier height and thickness by measuring the quasiparticle tunnelling current.

## 1. Introduction

Fabrication of high-Tc Josephson junctions suitable for investigation of macroscopic quantum phenomena is challenging. It has been theoretically predicted that the effects of dissipation due to nodal-quasiparticles on quantum dynamics are minimal for d0/d0 high-Tc Josephson junctions, where pair potential lobes of electrodes are aligned each other [1]. The d0/d0 configuration is realized in [100]-tilt grain-boundary junctions [2]. However, a tunnel barrier in the [100]-tilt $YBa_2Cu_3O_{7-x}$ (YBCO) grain-boundary junctions typically has a low resistivity and very high critical current density whereas opposite characteristics are favourable for the observation of the macroscopic quantum effects. A vacuum annealing [3, 4] or oxygen electromigration (EM) with a current applied across a grain boundary [5] are conventional ways to increase the resistance of the fabricated YBCO Josephson junctions. Unfortunately, an increase of the resistance with these techniques is accompanied by a significant decrease of a characteristic voltage $I_cR_n$, where $I_c$ is the fluctuation-free critical current and $R_n$ is the normal-state resistance of the Josephson junction. The deterioration of the $I_cR_n$-product occurs due to the reduction of the oxygen concentration and, hence, the decrease of an effective superconducting energy gap in regions close to the grain boundary. In this paper, we propose a new approach to the modification of the grain boundary properties when an electric field is applied along the grain boundary and perpendicular to the film surface. We use an insulator $BaTbO_3$ deposited atop of an YBCO film as a reservoir for the oxygen extracted from YBCO because of its high oxygen diffusion rate [6]. In addition, $BaTbO_3$ forms an atomically sharp interface with YBCO film and consists of elements compatible with YBCO [7, 8]. It is known that the oxygen diffusion in perovskites is enhanced in the grain boundaries or other local defects. When an electric field is applied along a grain boundary, the EM treatment has minimal effect on the oxygen concentration in the junction electrodes and the reduction of the order parameter in the regions neighbouring the grain boundary does not occur. Appling the electrical field directed parallel to the grain boundary, we increased the resistance of the [100]-tilt YBCO grain-boundary junctions from tens to several hundred Ohms without the reduction of the $I_cR_n$-product. The changes in the grain boundary caused by the fabrication process and the EM treatment were analysed using numerical calculations of the electrodynamics of a Josephson junction shunted by a capacitance.

## 2. Experimental details

A device used in this work consists of the [100]-tilt grain-boundary junction on a 2x12° $SrTiO_3$ (STO) bicrystal substrate and a $BaTbO_3$ layer with an electrode atop of the YBCO film. We fabricated the device in three steps. Figure 1 illustrates the fabrication steps and an electrical circuit used for the EM treatment. At the first step (figure 1a), a 70-nm-thick YBCO film was deposited on the bicrystal substrate by dc sputtering from a stoichiometric YBCO target at an oxygen pressure of 3.4 mbar. The substrate heater temperature was 935°C during the YBCO sputtering. After the YBCO film deposition, the substrate temperature was ramped down to 500°C and the film was annealed in $O_2$ (800 mbar) for 15 min at this temperature. The substrate was then cooled to room temperature and transferred to another sputtering machine where the YBCO film was covered by a 30 nm thick $BaTbO_3$ layer deposited by rf magnetron sputtering from a stoichiometric $BaTbO_3$ target at the pressure of 2 mbar through a shadow mask. The mask has an opening in the junction areas and keeps the contact area free of $BaTbO_3$. The YBCO film was exposed to air for several minutes during the transfer between the sputtering machines. The substrate heater temperature was 750°C during the $BaTbO_3$ sputtering. After the $BaTbO_3$ layer deposition, the substrate temperature was ramped down, the multilayer structure was annealed in $O_2$ (800 mbar) during 15 min at 500°C and cooled to room temperature. Then, it was patterned into four 1.8-μm-wide bridges crossing the grain boundary by a UV lithography and a wet etching with a Br/Ethanol etch [6]. Platinum contact pads were fabricated *ex situ* by a magnetron sputtering at room temperature. At the second step (figure 1b), a 150 nm thick $BaTbO_3$ layer was deposited atop of the grain-boundary junctions using the same procedure as for the 30-nm-thick $BaTbO_3$ layer. At the third step (figure 1c), platinum electrodes were fabricated above the grain-boundary junctions by dc magnetron sputtering through a shadow mask at room temperature. Photographs of the device with the top electrode and the bridge across the grain boundary are presented in figure 1d and figure 1e, respectively. After each fabrication step, current-voltage (*IV*) characteristics of the junctions were measured at temperatures of 4.2-4.3 and 77.6-77.9 K with a standard four-probe technique.

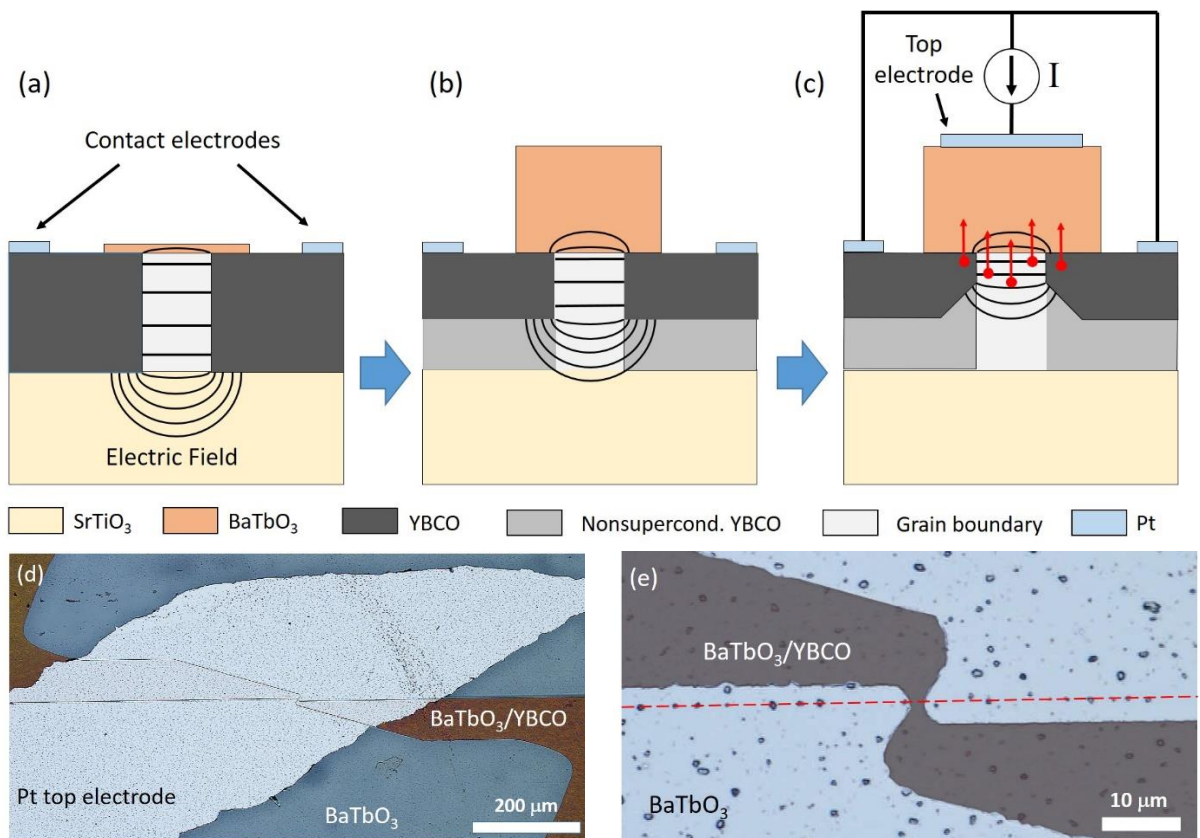

**Figure. 1**. (a-c) Fabrication steps and the electric field distribution in the vicinity of the grain boundary. Red arrows show the direction of the oxygen diffusion when the negative voltage is applied to the top electrode. (d) Photograph of the grain-boundary junction with current leads and the top electrode. (e) Photograph of the bridge across the grain boundary shown by a dashed red line.

EM treatments were performed at room temperature by applying a current between the top and contact electrodes, as shown in figure 1c. A Keithley 2614B sourcemeter was used as a current source.

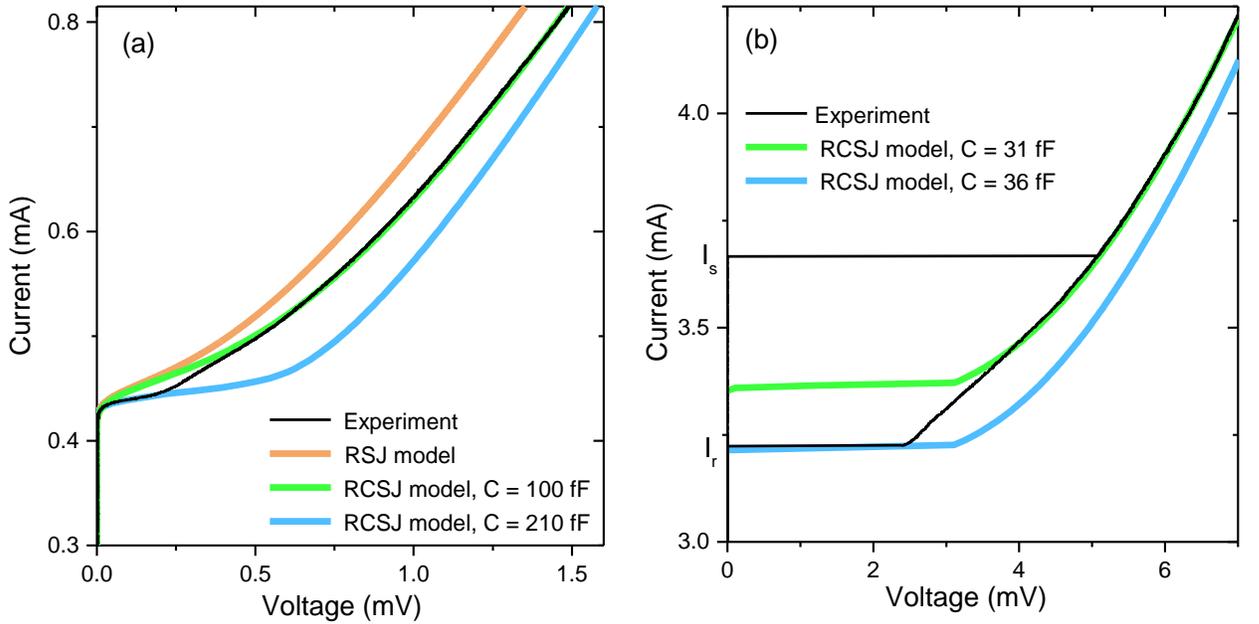

**Figure 2.** *IV* curves of the junction J1 at (a) $T = 77.9$ K and (b) $T = 4.2$ K (black lines) and the numerically calculated *IV* curves using the RSJ model (orange line) and the RCSJ model (blue and green lines).

## 3. Experimental results and discussion

As-fabricated junctions demonstrate continuous current-voltage characteristics at the temperature $T$ of 77.9 K and a voltage switching with a current hysteresis at $T = 4.3$ K. The representative *IV* curves of the as-fabricated junction are presented in figure 2 by black lines. The *IV* curves have significant deviations even at $T = 77.9$ K from the prediction of the Resistively Shunted Junction (RSJ) model, which is shown in figure 2a by an orange line. The experimental *IV* curve lies below the RSJ *IV* curve that is typical for a capacitive shunting of Josephson oscillations. Therefore, we employ the Resistively and Capacitively Shunted Junction (RCSJ) model to determine the junctions parameters. The method used for numerical calculations of the current-voltage characteristics of the Josephson junction in the presence of thermal fluctuation is described in Ref. [9]. Since the STO substrate has a frequency dependent dielectric constant, we use two different capacitance values at low and high voltages to fit the calculated *IV* curves to the experimental data. The results of the numerical calculations based on the RCSJ model are presented in figure 2 by green and blue lines. The normal-state resistance $R_n$ of the as-fabricated junctions was in the range of 2-3 Ω. At $T = 77.9$ K, we achieved good agreement between the experimental and calculated *IV* curves using the capacitance $C_1 = 50\text{-}100$ fF at the voltages above 0.5 mV (green line) and $C_2 = 100\text{-}200$ fF ar the voltages below 0.2 mV (blue line). At $T = 4.2$ K, we used the capacitance value $C_1 = 19\text{-}31$ fF to fit the results of the numerical calculations to the *IV* curve above a switching current and the capacitance $C_2 = 24\text{-}36$ fF to fit to the retrapping process (blue line). Similar capacitances of 20-28 fF can be calculated from the dependence of the Stewart-McCumber parameter $\beta_C = 2\pi I_c R_n^2 C/\Phi_0$ on the $I_r/I_s$ ratio at 4.2 K, where $\Phi_0$ is the magnetic flux quantum, $I_s$ is the switching current, and $I_r$ is the retrapping current [10]. The junction parameters obtained from the numerical calculations are presented in Table 1. We would like to note that the critical current density of the as-fabricated junctions is close to the critical current density of the YBCO film along the c-axis-tilt direction. Therefore, the magnetic vortices can enter into the film in the vicinity of the junction at currents slightly above the junction critical current resulting in asymmetric *IV* curves and a reduced critical current as it happened to the junctions J3 and J4 where we could not extract the capacitance values.

The calculated capacitances are well above an intrinsic grain-boundary capacitance, which can be estimated as $C_{gb} \approx \varepsilon_0 \varepsilon W d/t \approx 3\text{-}4$ fF, indicating that the substrate capacitance is a major contribution to the total capacitance of the as-fabricated junctions. Here $\varepsilon_0$ is the permittivity of free space, $\varepsilon$ is the grain boundary dielectric constant, $W = 1.8$ μm is the junction width, $d = 70$ nm is the YBCO film thickness, $t \approx 2$ nm is the average grain-boundary barrier thickness [11], and $\varepsilon/t = 2.5\text{-}3.3$ nm$^{-1}$ [12, 13]. The large contribution from the substrate capacitance suggests that the YBCO film is superconducting throughout the thickness and the ac electric field of the junction is coupled into the substrate, as shown in figure 1a. The observed increase of the junction capacitance with a decrease of the voltage across the junction, which is proportional to the frequency of the Josephson oscillations, and with an increase of the temperature is consistent with the temperature and frequency dependent behaviour of the STO dielectric constant [14].

Table 1. Junction parameters

|  |  | As-fabricated |  | After 150-nm-thick BaTbO$_3$ layer deposition |  | After top electrode fabrication | After low-current EM | After high-current EM |  |
|---|---|---|---|---|---|---|---|---|---|
|  |  | 77.9 K | 4.2 K | 77.9 K | 4.2 K | 77.6 K | 77.6 K | 77.6 K | 4.3 K |
| J1 | $R_n$ (Ω) | 2 | 1.94 | 5.15 | 4.97 | 9.4 | 9.4 | - | - |
|  | $I_c$ or $I_s$ (mA) | 0.439 | 3.67 | 0.142 | 1.03 | 0.0689 | 0.0707 | - | - |
|  | $C_1$ (fF) | 100 | 31 | 30 | 9.5 | 25 | 27 | - | - |
| J2 | $R_n$ (Ω) | 2.82 | 3.02 | 32.38 | 34.13 | 121.3 | 82.6 | 108.5 | 110.4 |
|  | $I_c$ or $I_s$ (mA) | 0.294 | 2.21 | 0.0156 | 0.202 | 0.00265 | 0.0051 | 0.00324 | 0.0575 |
|  | $C_1$ (fF) | 60 | 19 | 25 | 3.6 | 10 | 25 | 15 | 1.1 |
| J3 | $R_n$ (Ω) | 2.97 | 3.02 | 14.3 | 14.3 | 26.1 | 19.6 | 45.9 | 50.25 |
|  | $I_c$ or $I_s$ (mA) | 0.257 | 1.56 | 0.0387 | 0.436 | 0.0187 | 0.0264 | 0.00876 | 0.111 |
|  | $C_1$ (fF) | 50 | - | 35 | 9.5 | 16 | 24 | 15 | 2.5 |
| J4 | $R_n$ (Ω) | 2.15 | 2.37 | 44.2 | 43 | 143.7 | 120.1 | 327 | 355.3 |
|  | $I_c$ or $I_s$ (mA) | 0.383 | 2.51 | 0.0102 | 0.152 | 0.00186 | 0.00294 | 0.00045 | 0.0157 |
|  | $C_1$ (fF) | - | - | 25 | 3.4 | 10 | 10 | 5 | 0.32 |

To calculate the $I_cR_n$-product of the as-fabricated junctions, we used the current corresponding to the first maximum of the differential resistance $dV/dI$ at $T = 77.9$ K and the switching current $I_s$ at $T = 4.2$ K as the fluctuation-free critical current $I_c$. At 4.2 K, $kT/2E_J \approx 3 \cdot 10^{-5}$ for the as-fabricated junctions, and, hence, the values of the fluctuation-free critical current and switching current are very close to each other. Here, $E_J \approx \Phi_0 I_c/2\pi$ is the Josephson energy and $k$ is the Boltzmann constant. The $I_cR_n$-product of the as-fabricated junctions was in the range of 0.78-0.88 mV at 77.9 K and 4.7-7.1 mV at $T = 4.2$ K demonstrating a high quality of the junctions.

After the reference measurements, we deposited the 150-nm-thick BaTbO$_3$ layer atop of the junctions that resulted in a significant increase of the junction normal-state resistances to 5-43 Ω. The junctions with the normal-state resistance above 14.3 Ω demonstrated $IV$ curves with the voltage switching and current hysteresis already at $T = 77.9$ K, as shown in figure 3a. Since the switching to the resistive state at 77.9 K occurs at the same voltage for all three junctions, we attribute the hysteresis to an interaction of the Josephson oscillations with a resonance of an external system, whereas the hysteresis at $T = 4.2$ K, shown in figure 3b, is due to the capacitive shunting. Despite of the significant changes in the junction resistances, the $I_cR_n$-product of the junctions at $T = 4.2$ K was $I_cR_n = 5.4 – 7.0$ mV which is nearly the same as for the as-fabricated junctions. At $T = 77.9$ K, the junctions with the voltage switching show $I_sR_n$-product reduced to 0.45-0.56 mV. We suppose that the $I_sR_n$-product reduction is due to the premature switching in the resistive state by thermal fluctuations.

The effect of the premature switching on the junction critical current is usually estimated from the expression $I_s = I_c[1-[(kT/2E_J)\ln(\omega_p\Delta t/2\pi)]^{2/3}]$ (1), where $\omega_p = (2\pi I_c/\Phi_0 C)^{1/2}$ is the Josephson plasma frequency and $\Delta t$ is the time spent sweeping the bias current through the dense part of the $I_s(I)$ distribution [15]. However, using equation (1), we could not obtain reasonable $I_c$ values for $T = 77.9$ K where thermal fluctuations are strong. To evaluate the effect of the capacitance on the measured critical current in the presence of strong thermal fluctuations, we calculate a series of $IV$ curves for the Josephson junction with a capacitance in the range of 0-800 fF. The results of these calculations, shown in figure 4, clearly demonstrate the decrease of the measured critical current with increasing junction capacitance. Taking into account that the $I_cR_n$-product of the junctions was not changed at 4.2 K, we assume that it remained the same at 77.9 K as well. Therefore, we calculate the fluctuation-free critical current at $T = 77.9$ K required for the numerical simulations based on this assumption.

The capacitances obtained for the BaTbO$_3$ capped junctions were decreased compared to the original values. From the numerical calculations, we estimate the capacitance $C_1$ as 25-35 fF at 77.9 K and 3.4-9.5 fF at 4.3 K. We note that the capacitances $C_1$(4.3K) were close to the intrinsic grain-boundary capacitance $C_{gb}$. The decrease of the junction capacitance after the BaTbO$_3$ film deposition can be explained by an appearance of a nonsuperconducting YBCO layer between the substrate and the superconducting YBCO film, as shown in

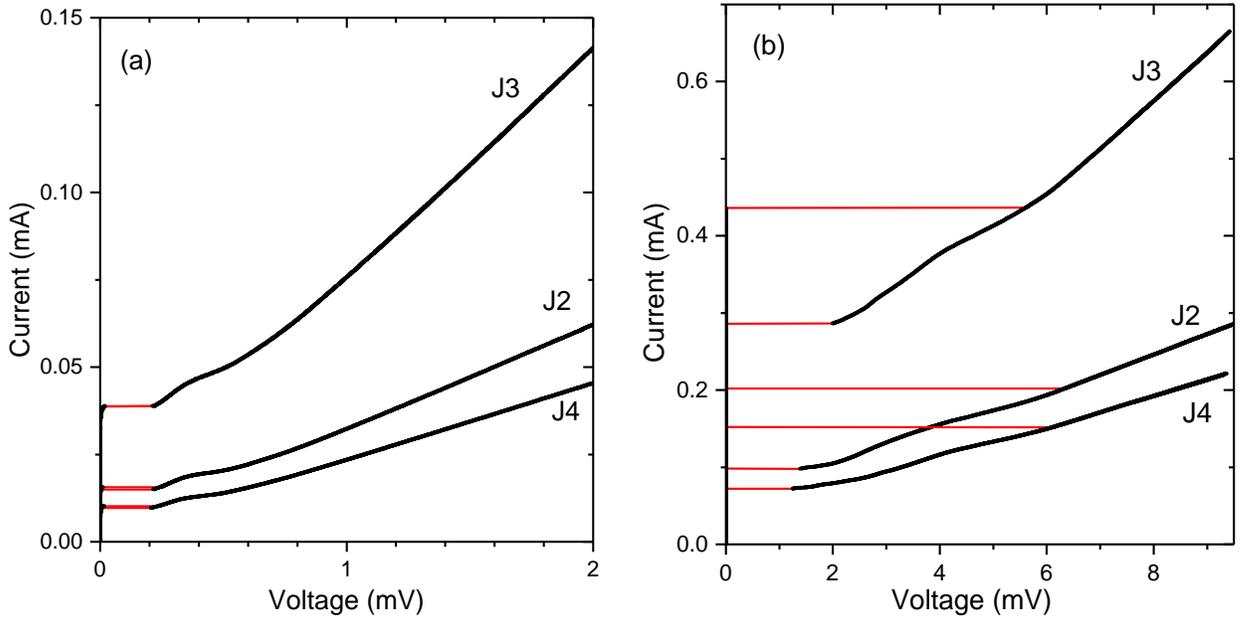

**Figure 3**. *IV* curves of the junctions J2-J4 at $T = 77.9$ K (a) and $T = 4.2$ K (b) after the 150-nm-thick $BaTbO_3$ layer deposition. Red lines indicate voltage switching.

figure 1b. The nonsuperconducting YBCO layer attenuates the coupling between the ac electric field of the junction and the substrate that results in the smaller total junction capacitance.

At the next step, we fabricated the Pt electrodes crossing the bridges with the grain-boundary junctions. The junction resistances increased up to 9.4-143.7 Ω after the top electrode fabrication, as shown in Table 1. The increase of the junction resistances cannot be explained by the heating effects because the power applied to a Pt target was low (3 W) and the changes in the junction or film parameters after the metal layer deposition have been never observed before. Therefore, we attribute the increase of the junction resistances to the electromigration of oxygen ions from the YBCO film to the $BaTbO_3$ layer. The electromigration is likely caused by an electric field gradient between the top and contact electrodes during dc magnetron sputtering. This assumption is confirmed by the corresponding decrease of the junction capacitances that results from the decrease of the superconducting film thickness in the vicinity of the grain-boundary.

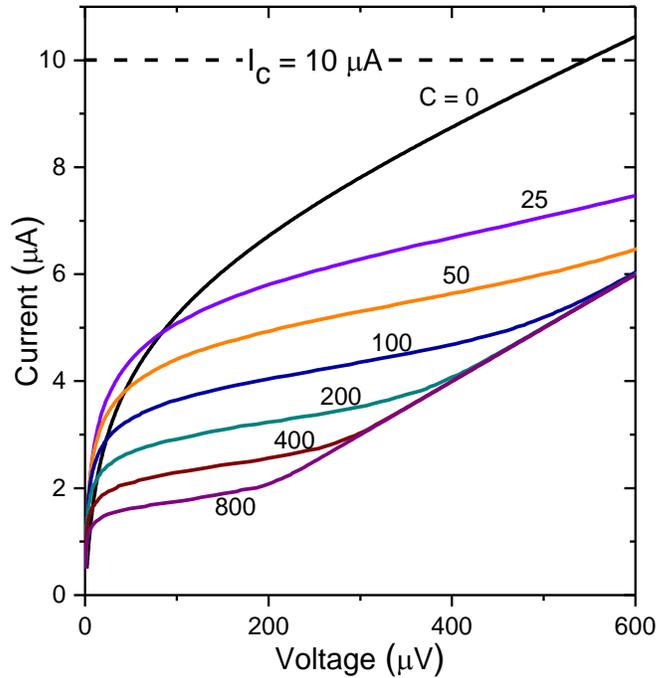

**Figure 4**. Numerically calculated *IV* curves of the capacitively shunted Josephson junction with $R_n = 100$ Ω and $I_c = 10$ μA at $T = 78$ K. The numbers near curves are the junction capacitance expressed in fF. Dashed line shows the value of the fluctuation-free critical current used in the calculations.

EM in YBCO is usually considered as a momentum transfer from moving holes in a ''hole wind'' to the oxygen atoms O1 in the Cu-O-Cu chains [5]. Here, we follow lattice site notation from Ref. [16]. Therefore, the oxygen atoms O1 have to diffuse towards a negative electrode. In conventional EM experiments with the YBCO grain-boundary junctions, when the low current is applied across the high-angle grain boundary, the grain-boundary resistance decreases and the critical current increases. This "healing" effect, which does not depend on a current direction, is explained by an oxygen ordering caused by a short-range redistribution of the oxygen in the Cu-O-Cu chains [5, 17]. When the current exceeds a certain threshold, the grain-boundary resistance starts to increase. The effects of an electromigration damage can be reversed by applying the smaller "healing" current in the opposite direction. The following mechanism is assumed for the resistance increase: at higher currents, the oxygen diffuse at a longer distance, and the oxygen diffusion stops at the high-angle grain boundary [18]. Therefore, the applied current create a highly doped region at the one side and the oxygen depleted region at the other side of the grain boundary that results in the increase of the resistance and decrease of the critical current. When the current is applied along the grain boundary, the dependence of the short- and long-range oxygen diffusion on the current has to be the same. However, the extended oxygen depleted regions in the vicinity of the grain boundary do not have to appear.

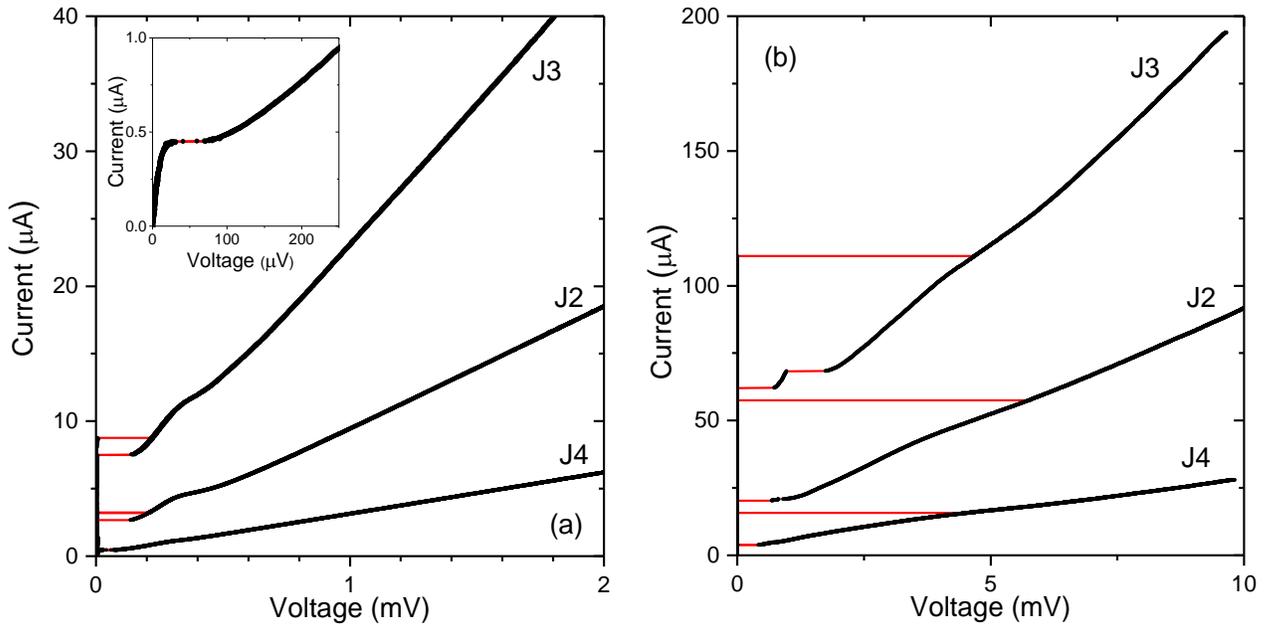

**Figure 5**. *IV* curves of the junctions J2-J4 at $T = 77.9$ K (a) and $T = 4.3$ K (b) after the high-current EM treatment. Inset shows an enlarged *IV* curve of the junction J4 at small bias currents. Red lines indicate voltage switching.

Following this model, we, first, applied the low current of 20 µA in different directions to the junctions J2-J4 during 5 min. No current was applied to the junction J1 that was used as a reference. The resistance of the junctions J2-J4 decreased by 15-30% demonstrating the "healing" effect, while the resistance of the reference junction J1 remained the same. Therefore, we conclude that the changes in the junction parameters are due to the applied current rather than a relaxation of the junction parameters because of the room temperature annealing. The capacitance $C_1(77.6$ K$)$ of the junctions slightly increased demonstrating that the thickness of the superconducting part of the film was increased in the vicinity of the grain boundary towards the substrate.

Then, we applied a negative voltage to the top electrode and increased the current up to 90-100 µA to give rise to the long-range oxygen diffusion in the direction of $BaTbO_3$. The duration of the high-current EM treatment was 15 min. The resistance of the junctions J2-J4 was increased by 33-200% up to 355 Ω. The *IV* curves of the junctions J2-J4 after the high-current EM treatment are shown in figure 5. The *IV* curve of the junction J4 with the normal-state resistance of 355 Ω does not show the superconducting part at $T = 77.6$ K because of the strong thermal fluctuations ($E_J < kT$). However, we observed a voltage jump at $I = 450$ nA, which is shown in the inset in figure 5a. The origin of this voltage switching is unclear because within the framework of the RCSJ model any sharp features at such small current have to be smeared out by thermal fluctuations and will be a subject of further study.

The capacitance $C_1(4.3$ K$)$ of the junctions J2-J4 was decreased to 0.32-2.5 fF that corresponds to the reduction of the Josephson junction thickness to 8-60 nm. The most pronounced decrease of the thickness of the superconducting part of the film has to occur in the regions close to the grain boundary where the rate of the oxygen diffusion is higher, as shown in figure 1c. The $I_sR_n$-product of 5.6-6.3 mV at 4.3 K is slightly lower than the $I_sR_n$-product of the as-fabricated junctions. However, if the effect of the premature switching in the resistive state is taken into account, the $I_cR_n$-product of the junctions after the EM treatment can be calculated as 6.0-7.0 mV which is the same as for the as-fabricated junctions.

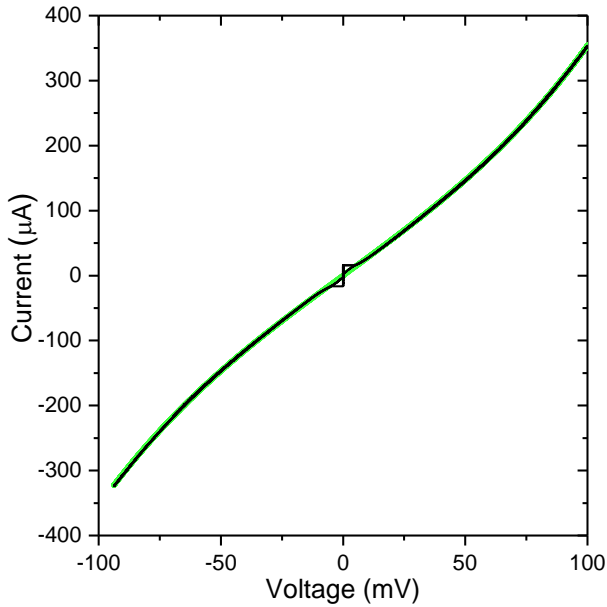

**Figure 6**. *IV* curve of the junction J4 at $T = 4.3$ K after the high-current EM treatment (black line). The green line shows the fit of a MIM tunnelling current to the experimental data.

A nominal current density through the YBCO/BaTbO$_3$ interface with an area determined by the overlap of the YBCO and Pt electrodes was 0.055 A/cm$^2$ for the low-current EM and 0.25-0.28 A/cm$^2$ for the high-current EM. These values are 6-7 orders of magnitude lower than the current densities (~1-2 MA/cm$^2$) used in the EM experiments with the current applied across the grain boundary [5, 19]. Assuming the same "hole wind" EM mechanism for our samples, we calculate the area where the high-current EM take place as 5000-10000 nm$^2$ that is roughly equal to the grain-boundary area $W \cdot t \approx 3600$ nm$^2$. Therefore, we conclude that the oxygen in our devices diffuses under applied current along the grain boundary and in regions of 1-2 nm from the grain boundary. We assume that the current-driven oxygen diffusion between YBCO and BaTbO$_3$ is possible because of the ability of Tb to be in both 3- and 4-valence state in the BaTbO$_3$. In the fully oxygenated BaTbO$_3$ Tb is in a 4-valence state while in partially oxygenated BaTbO$_3$ Tb is in a 3-valence state to keep a charge neutrality. It allows the slight oxygen deficiency, which favours oxygen diffusion.

The high-resistance junctions fabricated by the high-current EM treatment demonstrate both high $I_cR_n$-product and large current hysteresis favourable for the observation of the quantum effects at high temperatures. The crossover temperature between the macroscopic quantum tunnelling and thermal activation escape mechanisms can be estimated as $T_{cr} \approx \omega_p\hbar/2\pi k = 14\text{-}15$ K for the junctions J2-J4 [15].

In our calculation of the junction capacitance, we employed the barrier parameters reported for the [001]-tilt grain-boundaries. To obtain the parameters of the barrier of the [100]-tilt grain boundary we use the junction J4 where the heating effects at high voltages are negligible due to the high normal-state resistance. The *IV* curves of the junction J4 measured at $T = 4.3$ K up to 100 mV demonstrate excellent agreement with the $I = \beta(V+\gamma V^3)$ dependence typical for the Metal-Insulator-Metal (MIM) tunnel junctions, where $\beta$ and $\gamma$ are the constants depending on the barrier parameters. We extract an average barrier height $\varphi$ and thickness $t$ using an expression for the density of the tunnel current through a rectangular potential barrier in the MIM tunnel junction $J(V)=J_0[(\varphi-V/2)\cdot e^{-1.025\cdot t\cdot(\varphi-V/2)^{1/2}}-(\varphi+V/2)\cdot e^{-1.025\cdot t\cdot(\varphi+V/2)^{1/2}}]$ (2), where the current densities $J$ and $J_0$, barrier thickness t, and barrier height φ are expressed in A/cm$^2$, Å, and Volts, respectively [20]. Fitting the equation (2) to the experimental data, we obtain the average barrier thickness $t = 1.9311\pm0.0003$ nm and the average barrier height $\varphi = 80.93\pm0.14$ meV. The result of the numerical calculations is shown in figure 6 by the green line. Comparing the obtained barrier parameters with those reported for the 2x12° [001]-tilt grain-boundary junction, where $t = 2.1$ nm and $\varphi = 23$ meV [11], we note that the barrier height in the [100]-tilt grain-boundary junctions is significantly higher. We assume that the increased barrier height in the [100]-tilt grain-boundary junction is due to the absence of local states inside the barrier which are believed to be responsible for the relatively low $I_cR_n$-products of the [001]-tilt grain-boundary junctions. This assumption is in good agreement with our previous work where no local states inside the barrier were found for the [100]-tilt grain-boundary junctions [21].

## 4. Conclusions

We have fabricated the [100]-tilt grain-boundary YBCO Josephson junctions with the BaTbO$_3$ layer atop, which serves as the oxygen reservoir, and investigated the effects of the oxygen electromigration on the grain boundary for the electric field applied perpendicular to the YBCO film surface. As a result of the EM treatment, the normal-state resistance of the junctions was increased from tens to several hundred Ohms without the deterioration of the $I_cR_n$-product. We analysed the *IV* curves of the junctions before and after the electromigration treatment and conclude that the modification of the junction parameters occurs due to both decrease of the superconducting film thickness in the vicinity of the grain-boundary and changes in the tunnel barrier height and thickness. The average height and thickness of the grain-boundary barrier were extracted from the voltage dependence of the quasiparticle tunnelling current. The high-resistance junctions fabricated by the electromigration treatment could be promising for the investigation of the macroscopic quantum phenomena in high-Tc Josephson junctions.

**Acknowledgements**

M.L. is grateful to V.V. Pavlovskiy for valuable discussions and for providing software for the numerical simulation of the Josephson junction dynamics. M.L. and I.G. were partially supported within the framework of the state task AAAA-A19-119041990058-5.